# Nanofabrication of Plasmonic Circuits Containing Single Photon Sources

Hamidreza Siampour,* Shailesh Kumar, and Sergey I. Bozhevolnyi*

Centre for Nano Optics, University of Southern Denmark, Campusvej 55, DK-5230 Odense M, Denmark

**ABSTRACT:** Nanofabrication of photonic components based on dielectric-loaded surface plasmon-polariton waveguides (DLSPPWs) excited by single nitrogen vacancy (NV) centers in nanodiamonds is demonstrated. DLSPPW circuits are built around NV containing nanodiamonds, which are certified to be single-photon emitters, using electron-beam lithography of hydrogen silsesquioxane (HSQ) resist on silver-coated silicon substrates. A propagation length of 20±5 μm for the NV single-photon emission is measured with DLSPPWs. A 5-fold enhancement in the total decay rate and up to 63% coupling efficiency to the DLSPPW mode is achieved, indicating significant mode confinement. Finally, we demonstrate routing of single plasmons with DLSPPW-based directional couplers, revealing the potential of our approach for on-chip realization of quantum-optical networks.

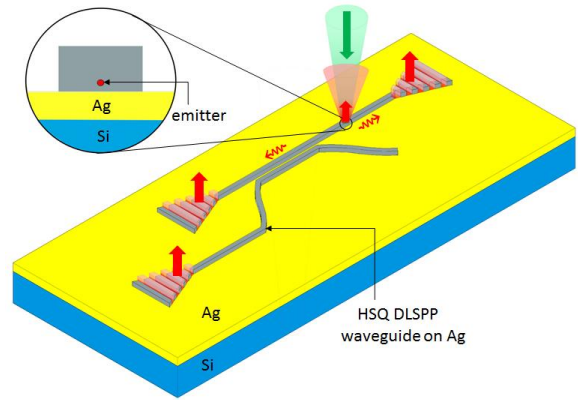

Strongly confined surface plasmon-polariton (SPP) modes can be efficiently excited by quantum emitters and the quantum information encrypted survives the photon-to-plasmon and plasmon-to-photon conversions.[1-5] The most important challenge in exploiting plasmonic circuitry for quantum-optical networks is related to inevitable SPP propagation loss by absorption (ohmic loss).[6-8] High transparency of dielectric waveguides facilitates information transport over long distances but the corresponding waveguide modes are diffraction-limited in their cross-sections.[6] Dielectric-loaded surface plasmon polariton (DLSPP) waveguides that confine SPPs laterally by using dielectric ridge waveguides patterned on a flat metal film support hybrid plasmonic-photonic modes and can thereby serve as a bridge between metallic nanoplasmonic components and dielectric waveguides.[9-12] Furthermore, DLSPP waveguides can be fabricated using standard lithography process,[13] as opposed to V-groove based plasmonic waveguides supporting channel plasmon polaritons that can have similarly low losses, but require advanced fabrication techniques, such as focused ion beam milling.[14-15] Furthermore, to accurately position quantum emitters in the vicinity of plasmonic structures or high-index dielectric waveguides,[16] scanning probe manipulation is often used.[14,17-18]

In this letter, we propose and experimentally demonstrate a top-down fabrication technique for deterministic positioning of DLSPP waveguides so as to incorporate nanodiamonds that are certified to contain a single nitrogen vacancy (NV) center, i.e., to be single-photon sources. NV centers are known to be stable and bright single-photon sources,[19-23] that also have an optically accessible electron and nuclear spin that can be used as qubits.[20] DLSPP waveguides are built with nanometer precision around pre-characterized nanodiamonds, whose locations are related to specifically designed and fabricated coordinate markers, using electron-beam lithography of hydrogen silsesquioxane (HSQ) resist deposited on silver-coated silicon substrates (Figure 1).

In the experiment, a silicon sample is coated with a silver film of 250 nm thickness, on which gold markers are made,

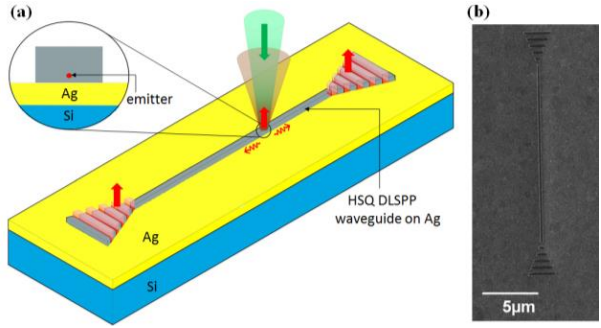

**Figure 1.** (a) Schematic of a DLSPP waveguide coupled to a nanodiamond containing single photon emitter. (b) Scanning electron micrograph of a HSQ waveguide, with 250 nm width and 180 nm height, fabricated on silver-coated silicon substrate (grating period = 550 nm).

and subsequently, nanodiamonds (Microdiamant MSY 0-0.05 micron GAF) are spin coated. The sample is then characterized by scanning in a fluorescence confocal microscope. A detailed description of the experimental setup is available in the Supporting Information (Figure S1). In Figure 2a, markers (+) and a preselected nanodiamond can be observed from the fluorescence image obtained from confocal microscopy. Lifetime, spectrum and autocorrelation measurements are taken for the nanodiamonds. HSQ e-beam resist (DOW CORNING XR-1541-006) is then spin coated (1200 rpm, 1 min) to make 180 nm film on Ag. Waveguide is fabricated using electron beam lithography onto the nanodiamond, which is found to be a single photon emitter. An AFM image of the fabricated waveguide is illustrated in Figure 2b. Post-fabrication CCD camera image shows the coupling of the emitter to the DLSPP waveguide, and subsequent emission from the gratings at the two ends (Figure 2c). In Figure 2d, antibunching dip observed in the second-order correlation function of the NV-center both before and after fabrication of the waveguide indicates a single photon emitter ($g^{(2)}(0) < 0.5$). The data is fitted with a model presented in ref. 24. Figure 2e shows the emission spectra of the NV-center taken before and after coupling, as well as the spectra of the out-coupled light from the grating ends A, and B. In Figure 2e, lifetime of the NV-center before and after coupling is presented. A lifetime reduction (from ~10 ns to ~6 ns) is observed for the coupled NV-center (Figure 2f). The lifetimes are obtained by the tail fitting of the measured data with a single exponential. First few ns of the data is avoided in the fitting, as it arises from background fluorescence. We have observed lifetime changes in the range 1.5 to 3.1, and on average the lifetime decreased by a factor of ~2.5 (see Figure S2 and S3 in the Supporting Information). On average the lifetime of NV-centers decreased by a factor of 2 due to a silver plane surface, from ~18 ns on fused silica substrate to ~9 ns on silver surface. This gives a ~5-fold enhancement in total decay rate ($\Gamma_{tot}$), which is in good agreement with the simulated $\Gamma_{pl}$ and $\Gamma_{tot}$, (discussed in detail later in this letter).

We measure propagation characteristics of the DLSSP waveguide for several waveguide samples of different lengths by comparing the attenuation of fluorescence signals $P_A$ and $P_B$ at the two ends of waveguides and their corresponding propagation distances $L_A$ and $L_B$ from coupled NV-centers. The 1/e propagation length, $L_P$, is extracted from the fluorescence signals at the two ends using $P_A/P_B = \exp[(L_A - L_B)/L_P]$, where we assume symmetric coupling in two directions, uniform losses across the waveguide and the same out-coupling efficiency at the grating ends. The collected data are fitted to obtain the propagation length of 20±5 μm for the DLSPP waveguide.

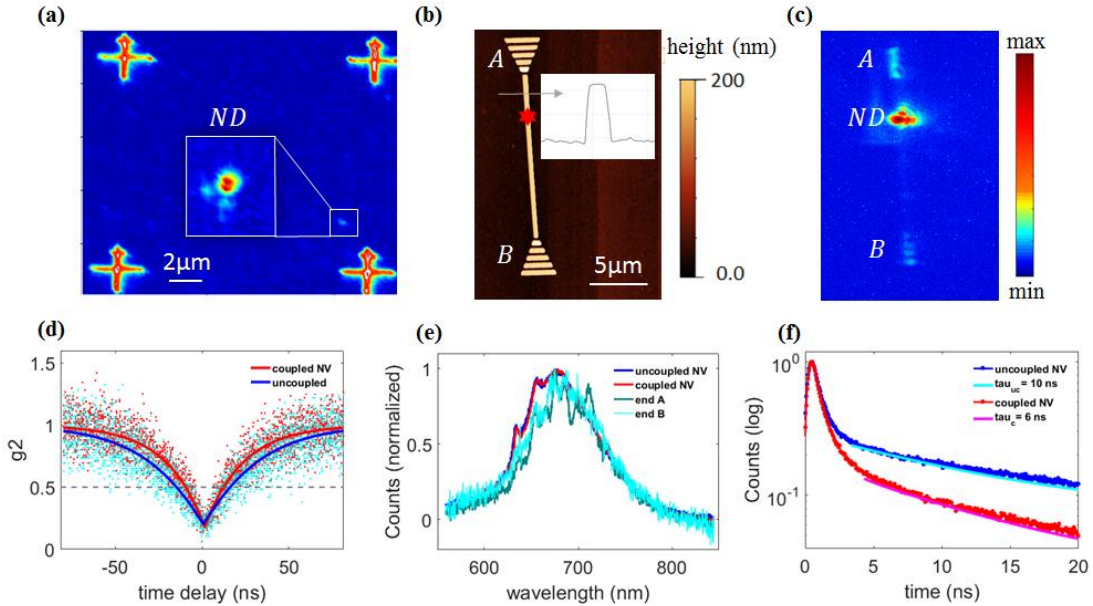

**Figure 2.** (a) Fluorescence scan map of nanodiamonds spin coated on a sample with markers of gold fabricated on silver surface. Structure indicates the nanodiamond and its position on the sample relative to the markers. (b) An atomic force microscope (AFM) image of the fabricated waveguide. The inset shows thickness profile across the grey arrow and indicates a 180 nm height for the waveguide. (c) Charge coupled device (CCD) camera image of the whole structure where the nanodiamond is excited and a fluorescence image of the focal plane is taken, is presented. Emission from the gratings at the ends of the waveguide, when nanodiamond is excited, confirms the coupling of NV-center to the waveguide mode. (d) Autocorrelation for the the NV-center before (blue) and after (red) coupling to the waveguide. (e) Spectrum taken from uncoupled NV (blue), coupled NV (red), and outcoupled light through the grating ends A (dark green) and B (light green) are presented (f) Lifetime of the NV-center before (blue) and after (red) coupling.

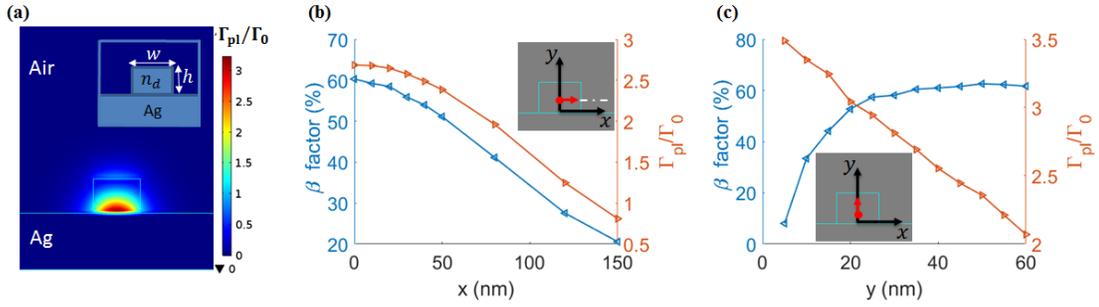

**Figure 3.** (a) Simulation result for plasmonic decay rate $\Gamma_{pl}/\Gamma_0$ shows a 3-fold emission enhancement at the optimal position of the emitter. The inset illustrates the DLSPP waveguide in the cross-section. Distance dependence of the plasmonic decay rate and spontaneous emission β factor for the DLSPP waveguide: (b) Emitter moved along dash-dotted white line outward from the middle of the waveguide, (c) Emitter moved along y-axis outward from the silver surface.

We calculate the emitter's decay rate into the plasmonic mode ($\Gamma_{pl}$) guided by the DLSPP waveguide using two dimensional (2D) finite-element modeling (FEM) method.[25] Normalized by the spontaneous emission decay rate in the vacuum ($\Gamma_0$), a 3-fold emission enhancement due to the plasmonic excitation is obtained (Figure 3a). Emission coupling efficiency (β factor) to the DLSSP waveguide is defined as the probability that the quantum emitter excites a single plasmonic mode. It can be described by the fraction of the emitted energy that is coupled to the plasmonic mode i.e. $\beta = \Gamma_{pl}/\Gamma_{tot}$, where $\Gamma_{pl}$ is the plasmonic decay rate obtained from the 2D simulation result shown in Figure 2a, and $\Gamma_{tot}$ is the total decay rate, including radiative decay rate, nonradiative decay rate, and the plasmonic decay rate of the quantum emitter coupled to the DLSPP waveguide.

A three dimensional (3D) FEM model with scattering boundaries surrounded the computational domain is implemented using COMSOL Multiphysics software and the total decay rate is extracted from the total power dissipation of the coupled emitter as explained in [25] and [26]. The coupling for a vertically oriented quantum emitter with a DLSPP waveguide is studied for two cases: (1) the distance of the emitter to the silver surface is fixed and equal to 35 nm, and the emitter is moved along dash-dotted white line outward from the middle of the waveguide as indicated with an arrow in Figure 3b-inset; and (2) the emitter is moved outward from the surface of the metal along y-axis (Figure 3c-inset). The distance dependence of the plasmonic decay rate ($\frac{\Gamma_{pl}}{\Gamma_0}$), and β factor for an emitter coupled to a DLSPP waveguide with 250 nm sidewidth, 180 nm height, and dielectric refractive index of $n_d$=1.41, are calculated for both cases in Figure 3b and Figure 3c, respectively. With optimized distance of the emitter to the silver surface, the β factor can reach 63%. We calculate the apparent radiative coupling factor, $\beta_r$, using $\beta_r \simeq \frac{[I_A+I_B]}{[I_A+I_B+I_{NV}]}$, where $I_A$ and $I_B$ are the emission intensity collected by galvanometric mirror scan at the ends of the waveguide and $I_{NV}$ is the emission intensity measured at the NV-center position (i.e. free space emission not coupled to the DLSPP mode). Here, we have assumed same collection efficiency from the NV-center as well as that from the ends, and have ignored the non-radiative decay, as the changes in observed lifetimes are not very high. An apparent coupling factor of 48% is found for the coupled DLSPP waveguide shown in Figure 2c. Taking into account the propagation length of the DLSSP waveguide, the corrected radiative coupling factor of 58% is achieved. This is in a good agreement with the simulation result in which the nonradiative emission is also taken into account and predicts a satisfactory precision in deterministic positioning of the DLSPP waveguide onto the nanodiamond. The ability of such coupled system to achieve efficient long-range energy transfer can be quantified by a figure of merit (FOM) defined as the product of the propagation length, Purcell factor ($\frac{\Gamma_{tot}}{\Gamma_0}$) and β-factor, normalized by the operation wavelength, as proposed in ref. 14. The proposed NV-DLSPP coupled system reaches a value of 83 ($\frac{\Gamma_{tot}}{\Gamma_0} = 5, \beta = 0.58, L_p$=20 μm, $\lambda$=0.7μm) for FOM, which is significantly larger than the value of 11.1 achieved for NV center coupled to V-groove waveguides presented in ref. 14.

Far-field characterization of the straight DLSPP waveguide using a super continuum source to excite one grating end and taking spectrum of the outcoupled light from another one is shown in Figure 4. In Figure 4a, one of the grating tapers is used for the DLSPP waveguide excitation, whereas the other one is used to monitor the transmission of the guided SPP mode through the structure. The transmission spectrum of the waveguide is taken and shown in Figure 4c (black line).

A directional coupler (DC) as a basic building block of plasmonic circuitry is designed based on DLSPP waveguides[27] and fabricated as shown in Figure 4b-inset. The DC consists of two rectangular DLSSP waveguides of height h=180 nm, width w=250 nm, refractive index $n_d$=1.41, the separation gap at the parallel section is g=200 nm, and the S-bends are based on cosine curves. The length of the parallel section (coupling length) is designed to impart a π phase shift at λ=700 nm between the symmetric and anti-symmetric DLSPP modes supported by the structure and calculated to be $L_c$=5.3 μm. Design rationale for the structure is explained in the Supporting Information (Figure S4). In Figure 4b, one of the grating tapers is used for SPP excitation, whereas the other ones are used to monitor the transmission of the guided DLSPP mode through the DC. The transmission spectra for both outcoupled ends together (B+C), and end C separately, are illustrated in Figure 4c. The spectrum of the end C shows a peak around λ=650 nm (instead of λ=700 nm that is designed by simulation) in which the corresponding coupling length, $L_c = \lambda/2(n_{eff}^+ - n_{eff}^-)$, imparts a π phase shift between the orthogonal uncoupled DLSPP modes with effective refractive indices of $n_{eff}^+$=1.23, and $n_{eff}^-$=1.16, respectively (see Figure S4 in the Supporting Information). The difference between experimental measurement and simulation results can be due to the difference in metal layer refractive index that are not exactly the same in fabrication as the ones used for simulation.

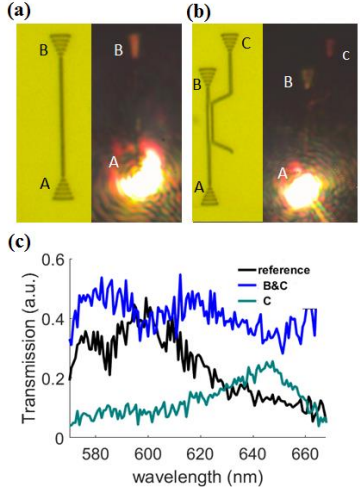
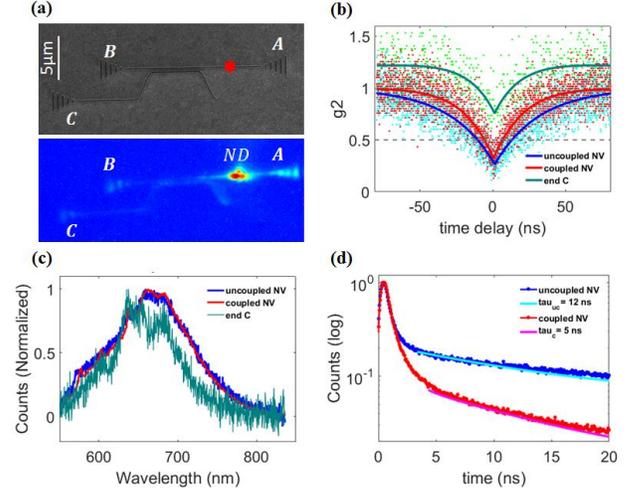

**Figure 4.** (a) Camera image of the straight DLSSP structure. In the inset, the microscopy image taken from the top DLSPP waveguides is illustrated. (b) Camera image of the DLSSP-based directional coupler. The microscopy image of the structure shows in the inset. (c) Transmission spectra taken for the straight DLSPP waveguide (black line), the both outcoupled ends B and C together (blue line), and the end C separately (green line).

**Figure 5.** (a) Charge-coupled device (CCD) camera image of the whole structure when the scanning laser beam is outcoupled through the grating ends (bottom), AFM image of the fabricated DLSPP-based DC (top). (b) Autocorrelation from the uncoupled NV (blue), coupled NV (red), and cross correlation taken between the coupled NV and outcoupled split end C (dark green). (c) Spectum taken from the coupled NV (red), outcoupled split end C (dark green). (d) Lifetime measured before (blue) and after (red) fabrication of the DC.

Following the same procedure for DLSPP waveguides, the DLSPP DC is fabricated onto a nanodiamond containing a single emitter. AFM image of the fabricated DC is presented in Figure 5a-top. Postfabrication fluorescent emission imaged on a charge coupled device (CCD) camera shows the coupling of the emitter to the structure, and subsequent emission from the grating at the three ends (Figure 5a-bottom). The antibunching dips in the second-order autocorrelation function of the coupled NV-center, and in the cross correlation between the coupled NV-center and the outcoupled end C are observed in the measurement results illustrated in Figure 5b. As indicated in Figure 5b, the antibunching dip for the cross-correlation is a little larger than 0.5 due to the worse signal to noise ratio at the waveguide end. In this experiment, it resulted in $g^{(2)}(0) > 0.5$.

For some other realizations, the dip corresponding to the cross correlation measurements remained below 0.5 as indicated in the Supporting Information (Figure S2-b and S5-b). Emission spectra of the coupled NV-center and outcoupled end C are shown in Figure 5c. Resonance peaks observed in the spectrum of the outcoupled end C can be described using $m\lambda_r = 2n_{eff}^{\pm}L_c$, where $\lambda_r$ denotes the resonance wavelength corresponding to the mth mode of the cavity length which is equal to the coupling length of the DC, $L_c$, and $n_{eff}^+$ and $n_{eff}^-$ are the effective refractive indices of the symmetric and anti-symmetric DLSPP modes propagating along parallel section of the DC. For the simulated parameters of $n_{eff}^+$ = 1.23, $n_{eff}^-$ =1.16, and $L_c$=5.3 μm, the corresponding resonance wavelengths are calculated to be close to 687 nm (m=19), and 653 nm (m=20), that fits well to the measured results in Figure 5c. Lifetime measurements of the NV-center, taken before and after coupling, show a reduction from 12 ns to 5 ns due to the emission enhancement of the coupled NV-center (Figure 5d). Additional experiment for DLSPP-based DC coupled to single NV-centers is available in the Supporting Information (Figure S5).

In conclusion, we have demonstrated DLSPP waveguide as a basic building block for plasmonic circuitry containing single photon sources. An emission coupling efficiency up to 63% and a 5-fold enhancement in total decay rate has been achieved due to strongly enhanced interaction between the single emitter and confined DLSSP mode. A large propagation length of 20±5μm has been observed experimentally that can be further enhanced by coupling of DLSSP waveguides into photonic waveguides. The DLSSP waveguides have been employed for making directional coupler, and can be also used for other components such as cavities to provide a platform for on chip processing of quantum information.

## ASSOCIATED CONTENT

### Supporting Information
Additional information is available in the supporting information.

## AUTHOR INFORMATION

### Corresponding Authors
* E-mail: hasa@mci.sdu.dk

* E-mail: seib@mci.sdu.dk

### Notes
The authors declare no competing financial interest.

## ACKNOWLEDGMENT
H.S., S.K., and S.I.B. acknowledge financial support from the European Research Council, Grant 341054 (PLAQNAP).

Supporting Information:

# Nanofabrication of plasmonic circuits containing single photon sources

Hamidreza Siampour,[*] Shailesh Kumar, and Sergey I. Bozhevolnyi[*]

Centre for Nano Optics, University of Southern Denmark, Campusvej 55, DK-5230 Odense M, Denmark

[*] E-mail: hasa@mci.sdu.dk
[*] E-mail: seib@mci.sdu.dk

# 1. Optical Set-up

The optical set-up used for characterization of the nitrogen vacancy (NV) centers as well as the NV-center dielectric loaded surface plasmon polaritons waveguide (DLSPPW) coupled system is presented in Fig. S1.

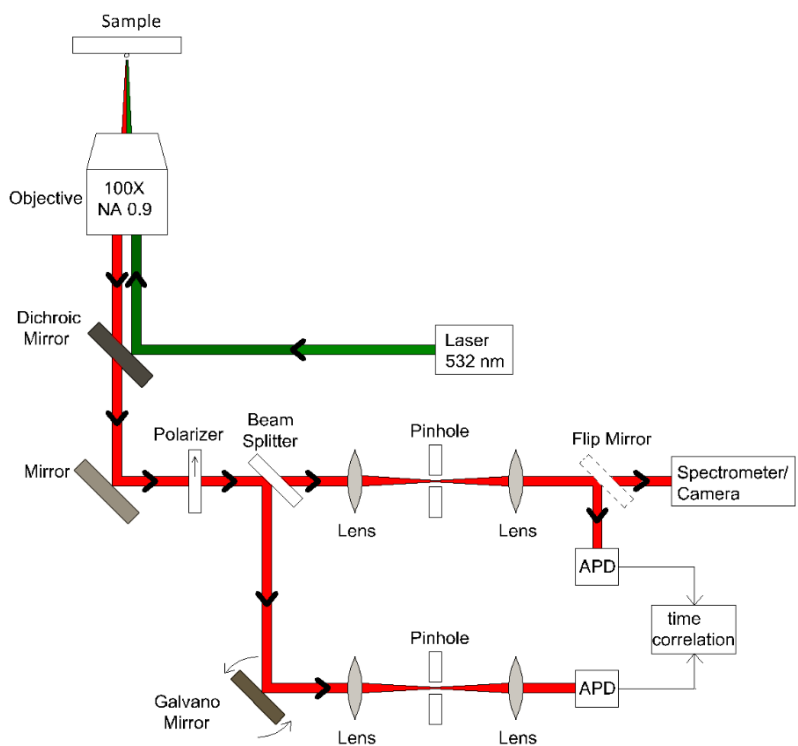

Figure S1. Schematic of experimental setup. Green line indicates excitation path from pump laser (532 nm) onto the sample, which is focused by a 100× (NA 0.90) objective. The fluorescence light, indicated by red line, is collected by the same objective, and passed through a dichroic mirror, polarizer and then beam splitter. When illuminated by a continuous wave laser, the emission from a single quantum emitter is split into two channels through the beam-splitter and then detected by two identical avalanche photodiodes (APDs) where we record time delay across the APDs to generate an intensity autocorrelation signal $g^{(2)}(t) = <I(t')I(t'-t)>$. Lifetime measurements are performed using pulsed excitation with pulse width/period of ~50 ps/400 ns. Postfabrication measurements are performed to show coupling of the emitter to the DLSPP waveguide where the nanodiamond is excited and a fluorescence image of the focal plane is taken either by a charge-coupled device (CCD) camera or a galvanometric mirror scan. Fluorescence spectrum of ND-waveguide system is taken by a grating spectrometer.

# 2. Excitation of DLSPP mode with a single NV-center

In Fig. S2 and S3, we present two NV-DLSPPW coupled systems. In Fig. S2, we present an $NV^-$ center coupled to a DLSPPW waveguide and in Fig. S3, a $NV^0$ center coupled to a DLSPPW waveguide is presented.

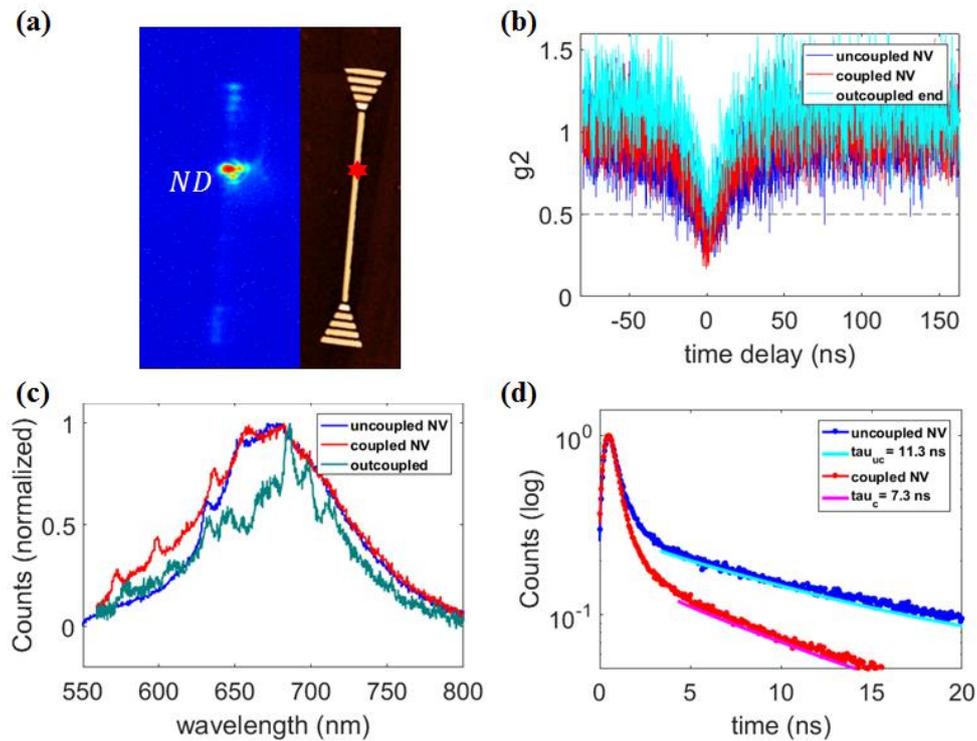

Figure S2. DLSPP waveguide coupled to a nanodiamond containing single nitrogen NV-center. (a) A charge-coupled device (CCD) camera image of the whole structure where the nanodiamond is excited and a fluorescence image of the focal plane is taken (left), and an atomic force microscope (AFM) image of the fabricated waveguide (right), are presented. The position of NV-center is indicated with a "star". Emission from the gratings at the ends of the waveguide, when nanodiamond is excited, confirms the coupling of NV-center to the waveguide mode. (b) Autocorrelation for the NV-center before (blue) and after (red) coupling to the waveguide, and cross correlation between the outcoupled grating end and NV-center (light green). (c) Spectrum taken from uncoupled NV (blue), coupled NV (red), and outcoupled light through the grating end (dark green) are presented. (d) Lifetime of the NV-center before (blue) and after (red) coupling.

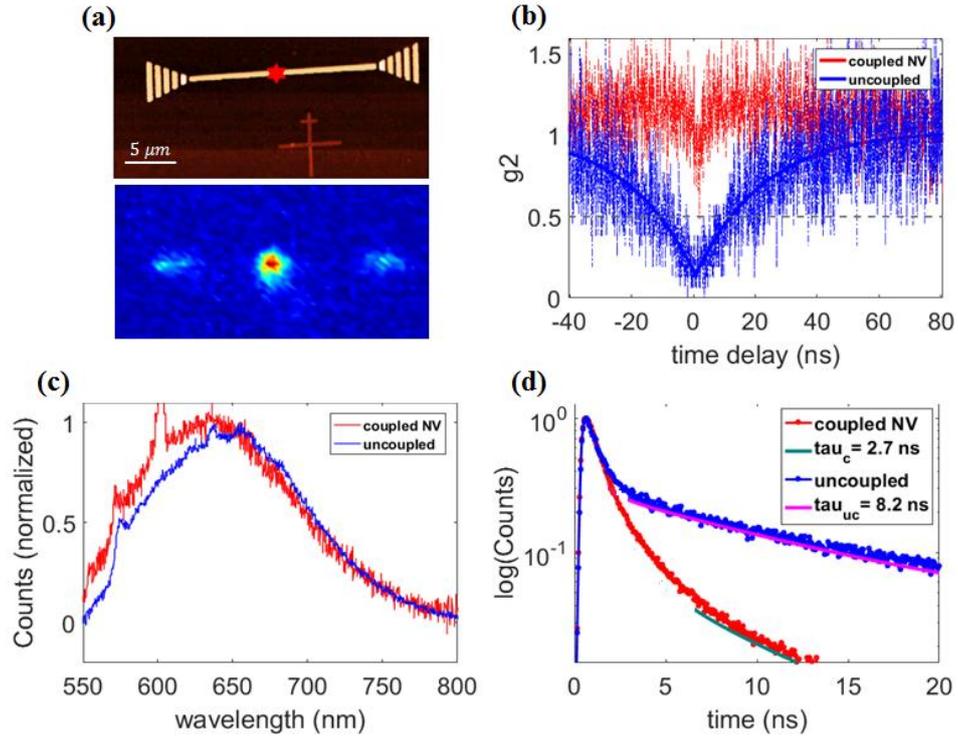

Figure S3. (a) An atomic force microscope (AFM) image of the fabricated waveguide (top), and a galvanometric mirror scan image (bottom), where the nanodiamond is excited and a fluorescence image of the focal plane is taken, is presented. Emission from the gratings at the ends of the waveguide, when nanodiamond is excited, confirms the coupling of NV-center to the waveguide mode. (b) Autocorrelation, (c) Spectrum, and (d) Lifetime of the NV-center before (blue lines) and after waveguide fabrication (red lines).

## 3. Routing of single plasmons in a DLSPPW based circuitry

When two DLSPPW waveguides are placed in the vicinity to each other, their individual modes cease to exist and two modes called antisymmetric and symmetric modes are supported, as presented in Fig. S4. The effective mode indices of the two modes are different and as the two modes propagate, the modes acquire a phase relative to the other mode. This results in a distribution of energy in the waveguides as a function of propagation distance, which can be used for routing of plasmon polaritons, in case of DLSPP, from one DLSPPW to other as explained in Fig. S4. The coupling length $L_c$ is given by, $L_c = \lambda/2(n_{eff}^+ - n_{eff}^-)$, where $n_{eff}^+$ and $n_{eff}^-$ are the effective mode indices of symmetric and antisymmetric modes, respectively, and $\lambda$ is the vacuum wavelength.

In Fig. S5, we present a system where emission from an NV-center is coupled to a DLSPPW and routed to another DLSPPW.

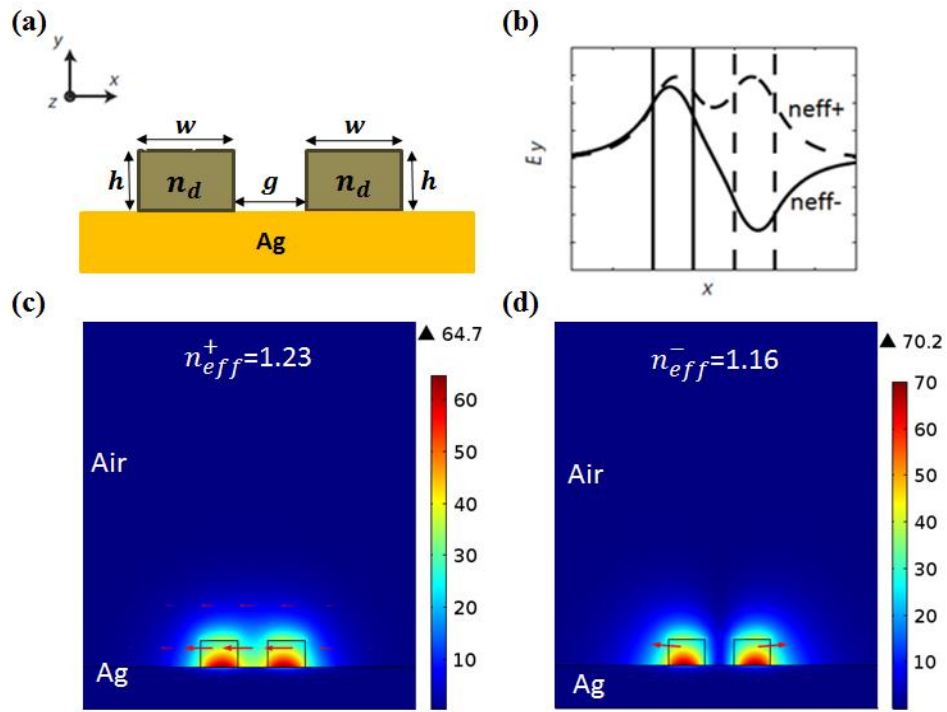

Figure S4. Simulation results of DLSPP-based directional coupler structure consists of two rectangular DLSSP waveguides of height h=180 nm, width w=250 nm, refractive index $n_d$=1.41, the separation gap at the parallel section is $g$=200 nm (a). The length of the parallel section (coupling length) to be $L_c$=5.3 μm in order to impart a π phase shift, at λ=700 nm, between the symmetric ($n_{eff}^+$) and anti-symmetric ($n_{eff}^-$) DLSPP modes (b) supported by the structure using $L_c = \lambda/2(n_{eff}^+ - n_{eff}^-)$. (c) Symmetric (d) Antisymmetric modes. Surface shows electric field norm (V/m) profile and red arrows indicate H-field.

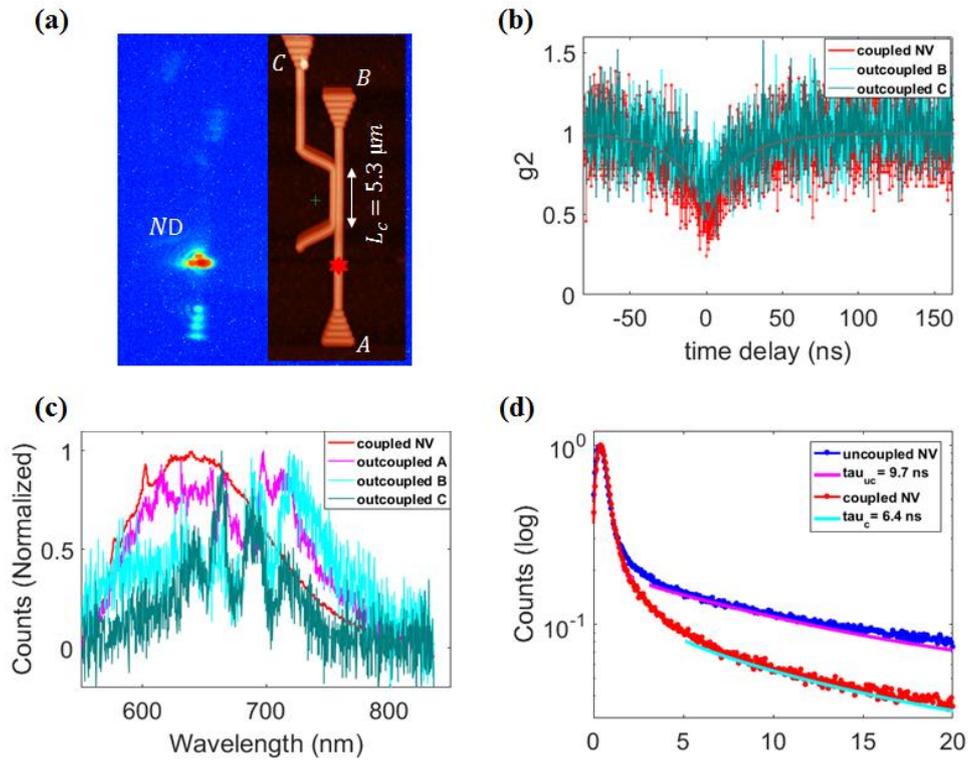

Figure S5. DLSPP-based DC coupled to a single NV-center in nanodiamond. (a) Charge-coupled device (CCD) camera image of the whole structure when the scanning laser beam is outcoupled through the grating ends (bottom), AFM image of the fabricated DC (top). (b) Autocorrelation from the coupled NV (red) and cross correlation taken between the coupled NV and outcoupled ends B (light green) and C (dark green). (c) Spectrum taken from the coupled NV (red), outcoupled ends B (light green) and C (dark green). The resonance wavelengths observed in the spectrum of the ends (B and C) are in good agreement with the ones corresponding to the coupling length, $L_c = 5.3 \ \mu m$, that are calculated to be 725 nm (m=18), 687 nm (m=19), 653 nm (m=20), and 622 nm (m=21). (d) Lifetime taken before (blue) and after (red) coupling to the DC.